\documentclass{article}         % for LaTeX 2e
\usepackage{amssymb}
\usepackage[dvips]{graphicx}
\begin{document}

\begin{flushright}
{\bf DFUB 2005-15}
\end{flushright}

\vspace{15mm}

\begin{center}
{\Large \bf Search for nuclearites with the SLIM detector}

\vspace{5mm}

S. Balestra$^1$, S. Cecchini$^{1,2}$, G. Giacomelli$^1$,  R. Giacomelli$^1$, M. Giorgini$^1$,
S. Manzoor$^{1,3}$, L. Patrizii$^1$, V. Popa$^{1,4}$ and O. Saavedra$^{5}$,
for the SLIM Collaboration

\end{center}
\vspace{3mm}

{\it 1. Dip. Fisica dell'Universit\`{a} di Bologna and INFN, 40127
Bologna, Italy \\
2. INAF/IASF, 40129 Bologna, Italy \\
3. PRD, PINSTECH, P.O. Nilore, Islamabad, Pakistan \\
4. Institute for Space Sciences, 77125 Bucharest - M\u{a}gurele, Romania \\
5. Dip. Fisica dell'Universit\`{a} di Torino and INFN, 10125
Torino, Italy }

\vspace{5mm}

{\bf Abstract.}
The strange quark matter (SQM) may be the ground state of QCD; nuggets of SQM could be
present in cosmic rays (CR). SLIM is a large area experiment, using CR39 and
Makrofol track etch detectors, presently deployed at the high altitude CR Laboratory of
Chacaltaya, Bolivia. We discuss the expected properties of SQM, from the point of view
of its search with SLIM. We present
also some preliminary results from SLIM.

\vspace{5mm}

\section{Introduction}
SLIM is a large area experiment (440 m$^2$) installed at the Chacaltaya CR lab
 since 2001; an additional 100 m$^2$ were installed at Koksil,
Pakistan, since 2003~\footnote{This paper refers to the Chacaltaya location only.}
\cite{slimnou,slimsimainou,slimslim,slimicrc}.
 With an
average exposure time of about 4 years, SLIM would be sensitive to a flux of
downgoing exotic particles
 at a level of  $10^{-15}$ cm$^{-2}$sr$^{-1}$s$^{-1}$. The main goal of SLIM is the
 search for intermediate and low mass magnetic monopoles,
 but it can be sensible to other exotica, as nuggets of SQM (known as ``nuclearites" or
 ``strangelets"), Q-balls, etc.

 We focus on the search for nuggets of SQM (both in the low mass region, in which
they are expected to behave more or less
like super-heavy nuclei, as well as in the intermediate
mass region, that is if they are heavy enough
to behave like micro-meteorites, but not so large to
be able to cross the Earth) in the  cosmic radiation, using the SLIM detector.
After a short description of the experiment and of the analysis procedures, we briefly
review the expected properties of SQM, relevant from the point of view of SLIM. We then
investigate the arrival and detection conditions of cosmic ray SQM nuggets in SLIM and
discuss some preliminary results yielded by the analysis of a part of the detector.

\section{The experiment}

SLIM (Search for Light and Intermediate mass Monopoles) is a large area detector (about 440 m$^2$)
presently exposed to penetrating CR at the Laboratory of Chacaltaya (Bolivia),
at an altitude of 5230 m a.s.l. The detector consists on modules of $24 \times 24$ cm$^2$,
made of 3 sheets of CR39 and Makrofol nuclear track detectors (NTDs),
and an aluminium absorber. This
stack structure is similar to that used by the MACRO track-etch subdetector
\cite{macrot,macroini,macro}.
Fig. 1 shows the sketch of one module.
\begin{figure}[t]
\begin{center}
\includegraphics[angle=0, width=0.8\textwidth]{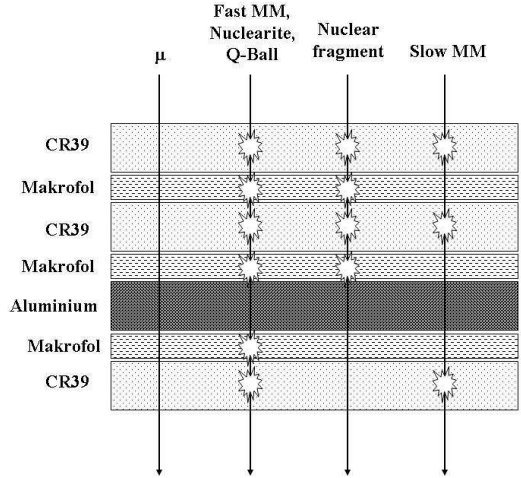}
\end{center}
\label{moduloslim}
\vspace{-2mm}
\caption{A sketch of a SLIM module. The stars indicate the detectors in which an etchable
track is produced by the passage of different particles.}
\end{figure}
Two additional Lexan sheets are placed on the top and the bottom of each stack; those detectors
are not used in our analysis, their role is only to
 absorb some of the $\alpha$ particles produced by
the Radon decays in the air around SLIM.
The response of NTDs
depends on the environmental conditions during their exposure; thus each module
is sealed in a mylar bag, filled with dry air at the normal atmospheric pressure (note that
the atmospheric pressure at Chacaltaya is 0.5 atm.).

The working principle of NTDs is qualitatively presented in Fig. 2.
\begin{figure}[t]
\begin{center}
\includegraphics[angle=0, width=0.7\textwidth]{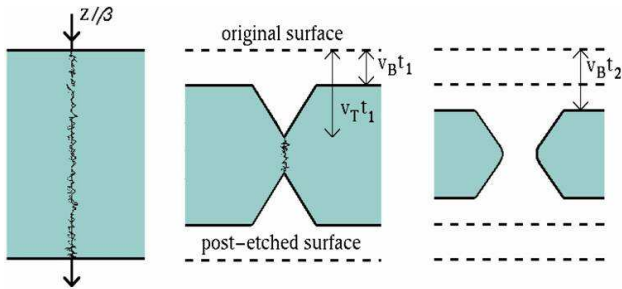}
\end{center}
\label{tracketch}
\vspace{-2mm}
\caption{Illustration of the working principle of track etch detectors. {\bf Left:}
the production
of a latent track, at the passage of a particle losing energy above the detection threshold.
{\bf Middle:} the etched cones obtained by chemical etching. {\bf Right:} The holes produced
after a longer etching.}
\end{figure}
When a particle crosses such a detector, if the restricted energy loss (REL) is above
some specific threshold, the polymeric structure is affected along is path yielding
the so called ``latent track". In the particular case of a particle of electric charge
$Z$ and velocity $\beta = v/c$, the produced damage is a function of $Z/\beta$.
The tracks became visible after chemical etching, typically in aqueous solutions of NaOH or
KOH, as the etching velocity along the latent track ($v_T$) is larger than the bulk
etching velocity
of the material $(v_B)$. In the initial stages of the etching, two cones are formed on both
sides of the detector. The geometry of those cones depends
on the REL of the incident particle;
the measurement of the base area or of the length of the cones allows, trough a suited
calibration, to determine it. If the etching is prolonged, the two cones
form a hole in the material; the REL information is lost,
but the detector becomes more appropriate for fast scanning operations.

NTDs may be calibrated using beams of relativistic ions. A typical set-up consists of
few sheets of detectors upstream of some target used for the beam fragmentation, and a set
of downstream sheets of NTDs that would record the tracks of the non-fragmented beam ions
as well as of the lighter ions produced in the target. Fig. 3 shows the distribution of
the averaged areas of the etch cones in CR39 (averages are made on only two detector faces)
produced by 158 AGeV In$^{49}$ and its fragments. The exposure was made at the CERN SPS.
\begin{figure}[t]
\begin{center}
\includegraphics[angle=0, width=0.9\textwidth]{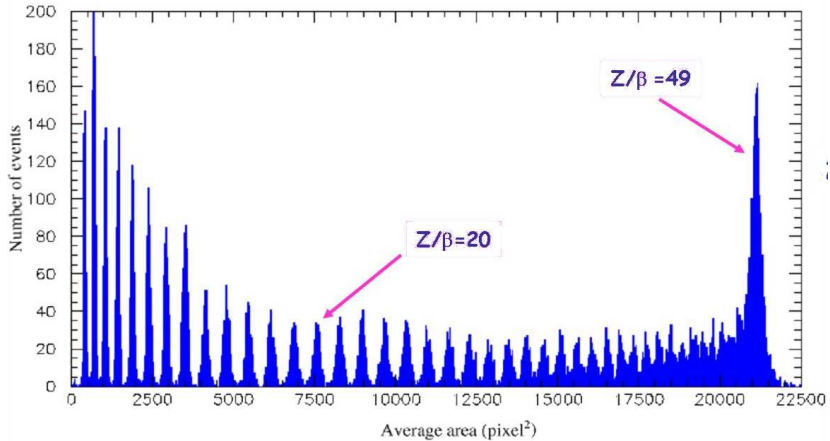}
\end{center}
\label{cali}
\vspace{-2mm}
\caption{Base-area distribution of the etch cones produced in the CR39 NTD (averaged over two
detector faces) produced by Indium ions at 158 AGeV and their fragments.}
\end{figure}
By computing the REL values coresponding to different relativistic ions, the calibration
curves for NTDs are obtained \cite{cal}. Such curves, for the CR39 (the squares) and for the
Makrofol\footnote{This is the first calibration of Makrofol based on the base-cone
areas \cite{manzoor}.}
(the circles) used in SLIM are presented in Fig. 4. The variable on the ordinate refers to
the so called ``reduced etching rate", $p = v_T/v_B$. The thresholds of the two detectors
 corespond to $p = 1$.
\begin{figure}[t]
\begin{center}
\includegraphics[angle=0, width=0.8\textwidth]{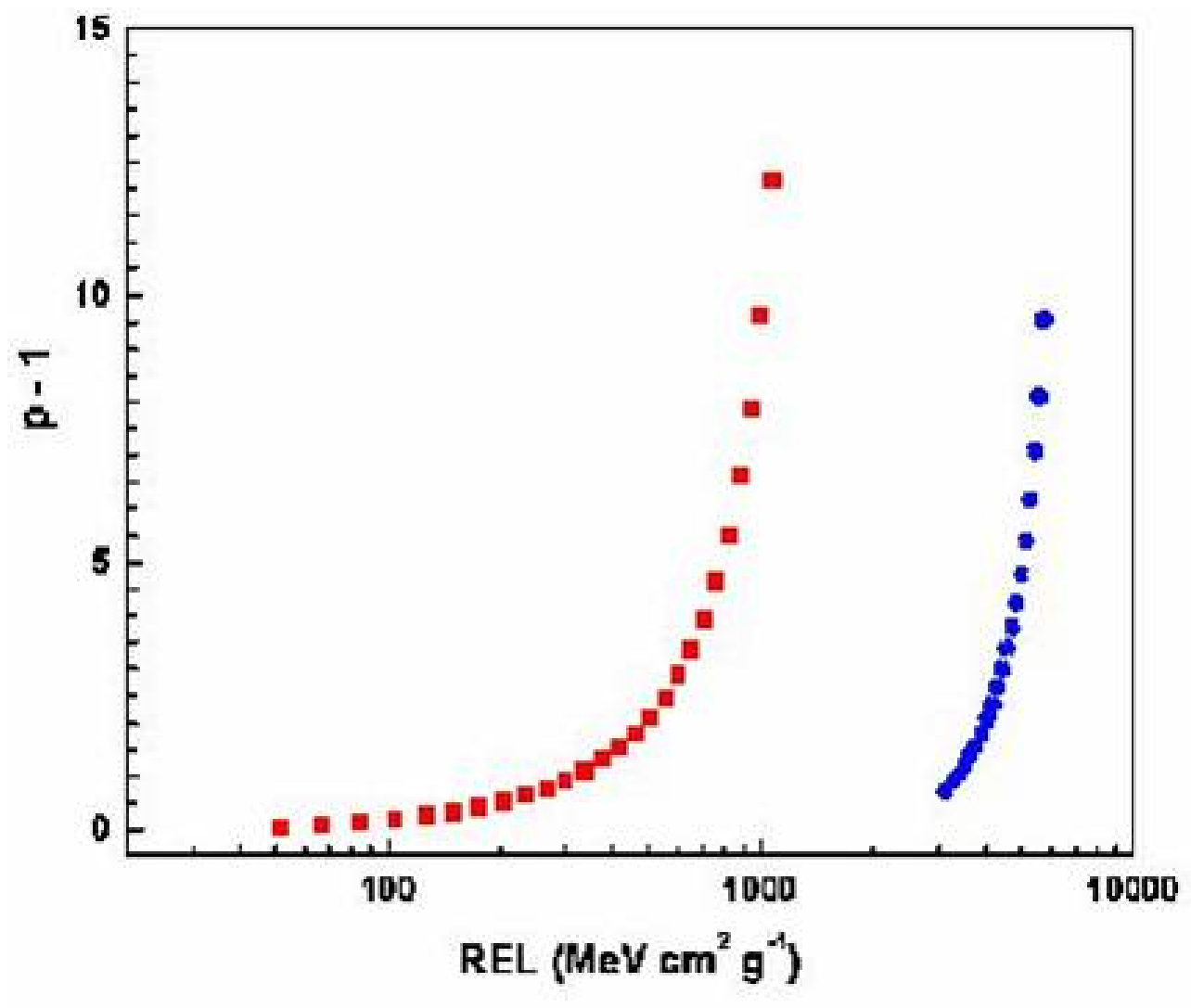}
\end{center}
\label{cali2}
\vspace{-2mm}
\caption{The callibration curves for the CR39 (the squares) and Makrofol (the circles) NTDs
used in SLIM.}
\end{figure}

The strategy to search for exotic candidates in SLIM is the following: firstly, we perform
a ``strong" chemical etching of the upper CR39 detector in each module. In those conditions,
large tracks (holes) are obtained, allowing an easy fast optical scan of the entire detector
surface. If tracks are found, the other two CR39 sheets are ``softly" etched (in order to
obtain measurable etched cones along the possible track) and scanned in the areas predicted by
the track in the first sheet. In the presence of tracks, the REL values and the orientation of
tracks are measured. In order to accept a candidate, double coincidences (between the detectors
above and bellow the Al absorber) are requested.
If this is the case, the Makrofol foils are etched and scanned too.
During the tests already performed on
part of the SLIM modules, no such coincidences were found.

\section{Strange Quark Matter in the cosmic rays: strangelets and nuclearites}

SQM could be the ground state of
quantum chromodinamics \cite{witten}. It is assumed that SQM is
made of
$u$, $d$ and $s$ quarks in nearly equal proportions. As the
chemical potential of the $s$ quarks in SQM is slightly larger than for
$u$ and $d$ quarks, SQM is always positively charged, so electrons  could
neutralize it. For small SQM nuggets ($M \lesssim 10^7$ GeV) the electrons would
form an electronic cloud around the quark core; for larger masses some electrons or,
for a quark bag radius $R \geq 1$\AA; $M \geq 8.4 \times 10^{14}$GeV \cite{ruhula}, all
the electrons
would be in equilibrium inside the SQM \cite{kasuya,calit}

 SQM is expected to have a density
slightly larger than ordinary nuclear matter \cite{witten,madsen}; the
relation between the mass $M$ of SQM lumps and their baryonic number $A$
would be
$M(\mbox{GeV}) \lesssim 0.93A$.

It was hypothesized
that nuggets of SQM, with masses from those of heavy nuclei (in this mass region we are
going to call them {\em strangelets})
to much higher values ({\em nuclearites}), produced in the Early Universe or
in violent astrophysical processes, could be present in the cosmic radiation \cite{ruhula}.

An upper limit for the flux of nuclearites may be obtained assuming
that they represent the main contribution to the local Dark Matter (DM) density,
$\rho_{DM} \simeq 10^{-24}$ g cm$^{-3}$ \cite{ruhula},
\begin{equation}
\Phi_{max} = \frac{\rho_{DM} v}{2 \pi M},
\label{dm}
\end{equation}
where $v$ and $M$ are the nuclearite average velocity and mass, respectively.

Calculations describing the production (through binary strange stars tidal disruption)
and the galactic propagation of cosmic ray nuclearites were recently published
\cite{mad}. The results could be valid  as
orders of magnitude for the entire mass
range of interest.

\subsection{Strangelets}

SQM  should be stable for all masses larger than about 300 GeV  \cite{ruhula}.
Strangelets with masses up to at least the multi-TeV region
could be ionized and could be accelerated to
relativistic velocities by the same astrophysical mechanisms of normal nuclei of the
primary CR.

They would interact with detectors (in particular NTDs) in ways similar to
heavy ions, but with different $Z/A$.
In ref. \cite{kasuya} SQM is described in analogy
with the liquid-drop model of normal nuclei; the obtained charge versus
mass relation is shown in Fig. 5A by the solid line, labeled ``(1)".
Other
authors found different relations:  $Z \simeq 0.1 A$
for $A \lesssim 700$  and $Z \simeq 8 A^{1/3}$ for larger
masses \cite{heis}: this charge to mass relation is shown in Fig. 5 %\ref{a-z}
as the dashed line, labeled ``(2)".
In \cite{madsen} it was assumed that quarks with different
color and flavor quantum numbers form Cooper pairs inside the SQM, the so-called
color-flavor locked (CFL) phase, increasing the stability of the strangelets.
In this case, the charge relation would be
$Z \simeq 0.3 A^{2/3}$, shown as the dash-dotted line in Fig. 5A labeled
``(3)".
\begin{figure}[t]
\begin{center}
\vspace{-55mm}
\includegraphics[angle=0, width=1\textwidth]{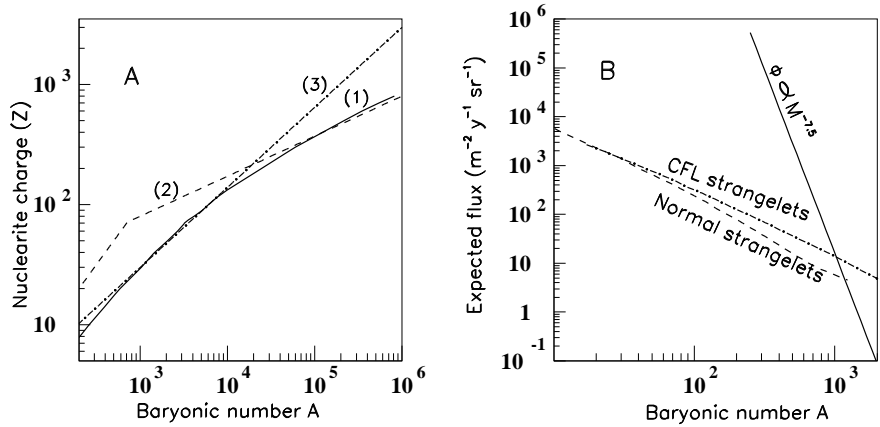}
\end{center}
\caption{{\bf A:} Strangelet electric charge versus mass for different hypotheses
discussed in refs. \cite{kasuya,madsen,heis}. See text for
details. {\bf B:} Expected fluxes for strangelets in the CR near the Earth.
The solid line corresponds to the assumption that their abundances
follow the same rule as heavy CR nuclei \cite{wilk1}. The dashed and dot-dashed
lines are from Ref. \cite{mad}, and refer to ``normal" and CFL strangelets. }
\label{strangelet}
\end{figure}

Several CR experiments reported possible candidate events that would suggest anomalously
low charge to mass (Z/A) ratios, which could correspond to those
expected for SQM \cite{kasuya}. Such candidates are
reviewed in \cite{wilk1,raha}. As strangelets
 could have the same origin as CR heavy nuclei, their
abundances in
the cosmic radiation could follow the same mass dependence, $\Phi
\propto M^{-7.5}$,  \cite{wilk1}. The existing candidates do
not contradict such an hypothesis.
The solid line in Fig. 6 %\ref{flux}
is the expected flux versus strangelet mass, assuming that the above assumptions
are correct.

Different nuclearite flux estimates were recently published \cite{mad}. They are
based on the hypothesis that large nuclearites (with masses $10^{-5} - 10^{-2}$
solar masses) are produced in binary strange stars systems, before their gravitational
collapse. The propagation inside the galaxy considers also the escape, spallation
(through which smaller nuclearites are produced) and  re-acceleration mechanisms.
Nuclearite decays are not considered, as SQM is supposed to be absolutely stable.
The predicted strangelet fluxes around the Earth are presented in Fig. 5B %\ref{flux}
for``normal" and CFL strangelets as the dashed and the dot-dashed lines, respectively.
The  differences  originate from the different charge-to-mass ratios.

The CR39 used in SLIM is sensitive (in the conditions of the ``strong" etching)
to particles with $\mbox{REL} \geq 200$ MeV g$^{-1}$ cm$^2$
\footnote{The threshold of a NTD depends on the etching conditions. A
relatively high threshold for CR39 was
chosen in order to reduce the background
due to recoil tracks, neutron interactions and the ambient radon radioactivity.}.
This implies a minimum $A$ (at the level of the detector) between 200 and 600, depending
on the chosen $A/Z$ model.

 If strangelets would interact with the Earth's atmosphere
in the same way as CR nuclei, they would not reach experiments at mountain altitude.
Different theoretical scenarios, both based on the SQM stability,
were introduced in order to allow their deep penetration
 in the atmosphere; none of those mechanisms would allow them anyway to reach
 sea level.

{\bf Mass and size decrease of strangelets during propagation.}
In \cite{wilk1} it was assumed that strangelets could  penetrate
the atmosphere if  their size and mass are reduced through successive interactions
with the atomic air nuclei.
This scenario is based on the spectator-participant picture.
 Two interaction models are considered: quark-quark
(``standard"), and collective (``tube-like").
 At each
interaction the strangelet mass is reduced by about the mass of a Nitrogen nucleus
(in the ``standard" model), or by more (in the ``tube-like model"),
while the spectator quarks form a lighter strangelet that continues its
flight with essentially the same velocity.
Once a critical mass is reached  ($A \simeq$ 300 - 400)
neutrons would start to evaporate from strangelets;  for $A \lesssim 230$
the SQM would become unstable and decay into normal matter.
In ref. \cite{wilk2} an estimate was made of the
sensitivity of a NTD experiment at Chacaltaya:
the mass number of a nuclearite penetrating the atmosphere
down to that altitude would be reduced by a factor $\simeq 1/7$.

In this scenario, the minimum strangelet $A$ value at the top of the atmosphere should be
between 1400 and 4200, in order to be detected in SLIM. Assuming different
strangelet structure and flux hypothesys, the expected fluxes in the Earth's vicinity
would range between some $10^{-12}$ and $10^{-15}$ cm$^{-2}$s$^{-1}$sr$^{-1}$.
More detailed calculations are given find in Ref. \cite{slimslim}.
Due to its
expected sensitivity, SLIM may discern between different combinations of flux - structure
hypotheses.

{\bf Accretion of neutrons and protons during propagation.}
A completely different propagation scenario was proposed in \cite{raha}. The
authors assume that strangelets would pick-up nuclear matter during
interactions with  air nuclei.
After each interaction, the strangelet mass would increase by about the atomic
mass of Nitrogen, with a corresponding  reduction of velocity. As the
mass grows larger, the loss in velocity becomes smaller.
They estimate that a
strangelet of an initial $A \simeq 64$ and an electric charge
of about +2 could arrive at about 3600 m a.s.l. with $A \simeq 340$ (3600 m is
the altitude of a proposed
NTD experiment in Sandakphu, India \cite{raha}).
This mechanism would also imply an increase of the electric charge of the strangelet,
 thus an increase of the Coulomb barrier; this may be its main
difficulty. The stability of low mass strangelets is
another questionable aspect of such a model; the expected fluxes would be larger than
in the fragmentation scenario, but, for this reason, they are hard to estimate. Such large
fluxes seem to be in disagreement with the preliminary SLIM results, presented in the
last Section of this paper.

\subsection{Nuclearites}

In \cite{ruhula} was postulated that elastic collisions
with atoms and molecules of the traversed medium
 are the only relevant energy loss
mechanism of non-relativistic
nuclearites with large masses,
\begin{equation}
\frac{dE}{dx}=-\sigma \rho v^2,
\label{ruhula1}
\end{equation}
where $\rho$ is the density of the traversed medium, $v$ is the
nuclearite velocity and $\sigma$ is its cross section:
\begin{equation}
\sigma= \left\{ \begin{array}{ll}
       \pi(3M/4 \pi \rho_N)^{2/3}  & \mbox{for $ M \geq 8.4 \times 10^{14}$ GeV
        (corresponding to $R_N \simeq 1$ \AA)} \\
       \pi \times 10^{-16} \mbox{cm}^2 & \mbox{for lower mass nuclearites}
       \end{array}
       \right. ,
\label{ruhula2}
\end{equation}
with $\rho_N = 3.6 \times 10^{16}$ g cm$^{-3}$.
 The cross section for nuclearites with $M < 8.4 \times 10^{14}$ GeV is
determined by their electronic cloud. For nuclearites gravitationally trapped in our Galaxy,
the average velocity would be $\beta = v/c \simeq 10^{-3}$.

Very large nuclearites ($M \gtrsim 3 \times 10^{22}$ GeV) could cross the Earth; as the
SLIM sensitivity is above the DM flux limit (Eq. \ref{dm}) we consider here only down-going
nuclearites.

A nuclearite of mass $M$
entering  the atmosphere with an initial
velocity $v_0 \ll c$, after crossing a depth $L$ will be slowed down to
\begin{equation}
v(L) = v_0 e^{-\frac{\sigma}{M}\int_0^L{\rho(x) dx}}
\label{vit}
\end{equation}
where $\rho(x)$ is the air density.
In the following we  consider the parametrization of the standard atmosphere
from \cite{shibata},
\begin{equation}
\rho(h) = a e^{-\frac{h}{b}} = a e^{-\frac{H-L}{b}},
\end{equation}
where $h$ is the altitude,
$a = 1.2 \times 10^{-3}$g cm$^{-3}$ and $b \simeq 8.57 \times 10^5$ cm; $H$
is the total
height of the atmosphere ( $\simeq$ 50 km).
The integral in Eq. \ref{vit} may be solved analytically.

Fig. 6A
 shows the velocity with which nuclearites of different masses reach heights
 corresponding to typical balloon experiments
 ($\simeq 40$ km),
to SLIM, (5.29 km)
and sea level. A
computation valid for  MACRO \cite{macro} (at a depth of 3400 mwe) is also included.
The velocity thresholds for detection
in CR39
 and in Makrofol
 are shown as the dashed curves.
\begin{figure}[t]
\begin{center}
\vspace{-55mm}
\includegraphics[angle=0, width=1\textwidth]{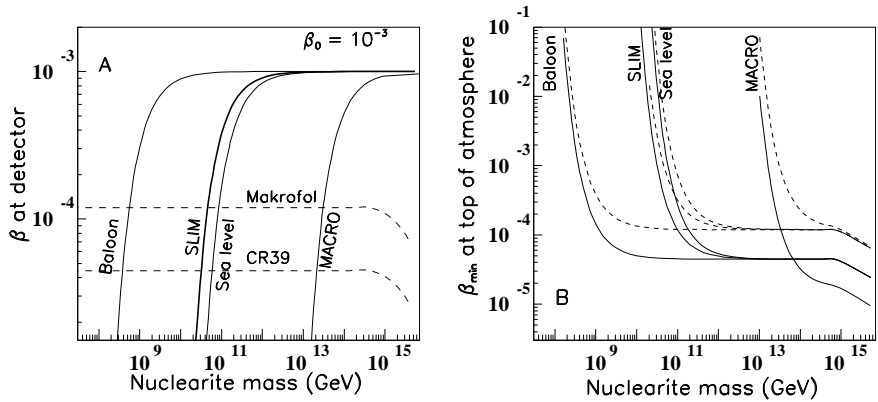}
\end{center}
\caption{{\bf A:} Solid lines: arrival velocities of IMNs at different depths
versus nuclearite mass, assuming an
initial velocity outside the atmosphere of $\beta = 10^{-3}$.
The nuclearites are supposed to come from above, close to the vertical direction.
The dashed lines
show the detection thresholds in CR39 (in the SLIM etching and Makrofol. {\bf B:}
Nuclearite detection conditions in CR39 (solid curves) and Makrofol (dashed curves),
for experiments located
at different altitudes. }
\label{nucleariti}
\end{figure}
The decrease of the velocity thresholds for nuclearite
masses larger than $8.4 \times
10^{14}$ GeV is due to the
change in the nuclearite cross section, according to Eq.
\ref{ruhula2}.
An experiment at the Chacaltaya altitude lowers the minimum detectable
nuclearite mass by a factor of about 2 with respect to an experiment
performed at sea level.
If the mass abundance of nuclearites decreases strongly with increasing
mass
this could yield an
important increase in sensitivity.

More general nuclearite detection conditions in CR39 and Makrofol
(expressed as the minimum velocity at the top of the atmosphere
versus the nuclearite mass) for different experimental
locations are shown in Fig. 6B. In this case, the constraint
 is that nuclearites have the minimum velocity at the
detector level in order to produce a track\footnote{The CR39 threshold in the MACRO
case was lower than for SLIM, due to the very low background at Gran Sasso Lab.}.

Searches for nuclearites (mostly IMNs) were performed by different experiments
\cite{naka,orito}. The  best flux upper limit was set by the MACRO
experiment:  for nuclearites
with $\beta \simeq 10^{-3}$ and $10^{14}$ GeV $<M< 10^{22}$ GeV,
the 90\% C.L. upper limit is
at the level of $2 \times 10^{-16}$ cm$^{-2}$sr$^{-1}$s$^{-1}$ ,
 as a byproduct of the search
for GUT magnetic monopoles \cite{macromono,macrogg}.

\section{Preliminary results and conclusions}
Till now, we analyzed about 214 m$^2$ of the SLIM detector, with an averaged exposure time
at Chacaltaya of 3.5 years. As no candidate was found, the present 90\% CL upper limit for a
flux of downgoing strangelets and nuclearites
with $M \gtrsim 3 \times 10^{13}$ GeV, valid also for relativistic monopoles, is
$3 \times 10^{-15}$ cm$^{-2}$s$^{-1}$sr$^{-1}$. This limit disfavors the hypothesis
of the accretion of matter by strangelets going down in the atmosphere, and can constrain some
production or propagation models.

 We intend to complete the
analysis by the end of 2006. Even if no magnetic monopole or SQM candidate will be found,
SLIM will yield significant limits in mass regions not yet explored by other experiments,
and will impose strong constraints on different scenarios describing the production and
propagation of strangelets.

\bigskip
{\small We aknowledge many useful discussions with other members of the SLIM Collaboration.
Special thanks are due to the technical staff of the NTD lab. of INFN Bologna.}

\end{document}